\documentclass[aps,prl,twocolumn,floatfix, showpacs]{revtex4}
\usepackage{bm,dcolumn,amsmath,graphicx}
\begin{document}
\title{Highly-charged ions for atomic clocks, quantum information, and\\ search for $\alpha$-variation}
\author{M. S. Safronova$^{1,2}$}
\author{V. A. Dzuba$^{3}$}
\author{V. V. Flambaum$^{3}$}
\author{U. I. Safronova$^{4,5}$}
\author{S. G. Porsev$^{1,6}$}
\author{M. G. Kozlov$^{6,7}$}

\affiliation {$^1$University of Delaware, Newark, Delaware, USA}
\affiliation {$^2$Joint Quantum Institute, NIST and the University of Maryland, College Park, Maryland, USA}
\affiliation {$^3$The University of New South Wales, Sydney, Australia}
\affiliation {$^4$University of Nevada, Reno, Nevada, USA}
\affiliation {$^5$University of Notre Dame, Notre Dame, Indiana, USA}
\affiliation {$^6$Petersburg Nuclear Physics Institute, Gatchina,  Russia}
\affiliation {$^7$St.\ Petersburg Electrotechnical University ``LETI'', St.\ Petersburg, Russia}
\date{\today}

\begin{abstract}
We propose 10 highly-charged ions as candidates for the development of next generation atomic clocks, quantum information, and search for $\alpha$-variation.
They have long-lived metastable states with transition frequencies to the ground state between 170-3000~nm, relatively simple electronic structure, stable isotopes and high sensitivity to $\alpha$-variation (e.g., Sm$^{14+}$, Pr$^{10+}$, Sm$^{13+}$, Nd$^{10+}$). We predict their properties crucial for the experimental exploration and highlight particularly attractive systems for these applications.
\end{abstract}
 \maketitle

The past few years marked unprecedented improvements in both the accuracy and stability of optical frequency standards~\cite{1,2,3,Ybclock,MadDubZho12}. Sr lattice clock group has just reported the achievement of the $6.4\times10^{-18}$ accuracy \cite{3}.
This remarkable progress poses the question of what are the novel schemes for the next generation of clock development that may achieve the accuracy at the next decimal point, $10^{-19}$. This work proposes 10 highly-charged ions (HCI) as candidates for the development of next generation atomic clocks and other applications. Further development of even more precise frequency standards is essential for new tests of fundamental physics, search for the variation of fundamental constants, very-long-baseline interferometry for telescope array synchronization, and development of extremely sensitive quantum-based tools for geodesy, hydrology, and climate change studies, inertial navigation, and tracking of deep-space probes \cite{2,3}.

The modern theories directed toward unifying gravitation with the three other fundamental interactions suggest variation of the fundamental constants in an expanding universe~\cite{Uza03}. Studies of the quasar absorption spectra indicate that the fine-structure constant $\alpha=e^2/\hbar c$ may vary on a cosmological space-time scale \cite{MurWebFla03,WebKinMur11}. This result has not been confirmed yet by other groups (see \cite{RSGP12} and references therein). As a result, the status of the observational search for $\alpha$-variation is presently unclear and further investigations are required \cite{BerFla11}.
Spatial $\alpha$-variation hypothesis can be tested in terrestrial studies if sensitivity $\delta \alpha/\alpha \sim 10^{-19} \ {\rm yr}^{-1}$ is achieved~\cite{BerFla2012}. Current best limit of $\delta \alpha/\alpha \sim 10^{-17} \ {\rm yr}^{-1}$ comes from measuring the frequency ratio of Al$^+$/Hg$^+$ clocks~\cite{RosHumSch08}.

The signature feature of the system suitable for  atomic clock development and subsequent tests of the $\alpha$-variation is the availability of the optical (or  near-optical) transition between a ground state and a long-lived metastable state.  It appears from the first glance that this requirement
reduces atomic clocks based on HCI to transitions between states of the ground state configuration (see, e.g.~\cite{DDf12,DerDzuFla12}) which are not sensitive to variation of $\alpha$.
However, this turns out not to be the case. Despite very large ionization energies, certain ions have transitions belonging to different configurations that lie in the optical range due to level crossings and strongly enhanced
sensitivity to variation of $\alpha$ ~\cite{BerDzuFla10,BerDzuFla11b}.
Highly-charged ions are less sensitive to external perturbations than either neutral atoms or singly-charged ions due to their more compact size. Recent studies of uncertainties \cite{DerDzuFla12,DzuDerFla12} have shown that the fractional accuracy of the transition frequency in the clocks based on HCI can be smaller than 10$^{-19}$.

One of the main obstacles for the experimental work toward the development of highly-charged ions for these applications is the lack of experimental data for these HCI systems as well as difficulties in accurate  theoretical predictions of the transition wavelengths owing to severe cancellations of energies of upper and lower states (since we are interested to work near level crossing). The goal of the present work is to resolve this problem. We identified \textit{all of the highly-charged ions}  that are particularly well suited for the experimental exploration, i.e. satisfy the following criteria: 1) existence of long-lived metastable states with transition frequencies to the ground states ranging between 170-3000~nm, 2) high sensitivity to $\alpha$-variation, and 3) existence of stable isotopes. Other practical consideration include relatively simple electronic structure, i.e. with one, two, or three valence electrons above the closed core, which excludes  core-excited states  such as considered in ~\cite{BerDzuFla11b}.  We find that  only the ions in four isoelectronic sequences, Ag-like, Cd-like, In-like, and Sn-like satisfy these criteria. There are no useful level crossings for isoelectronic sequences corresponding to the previous row of the periodic table ~\cite{BerDzuFla12}. The level crossings for the isoelectronic sequences in the next row of the periodic table that includes Au-like, Hg-like, Tl-like, and Hg-like ions occur for ions with no stable isotopes ~\cite{BerDzuFla12}.

\begin{figure}
  \includegraphics[scale=0.45]{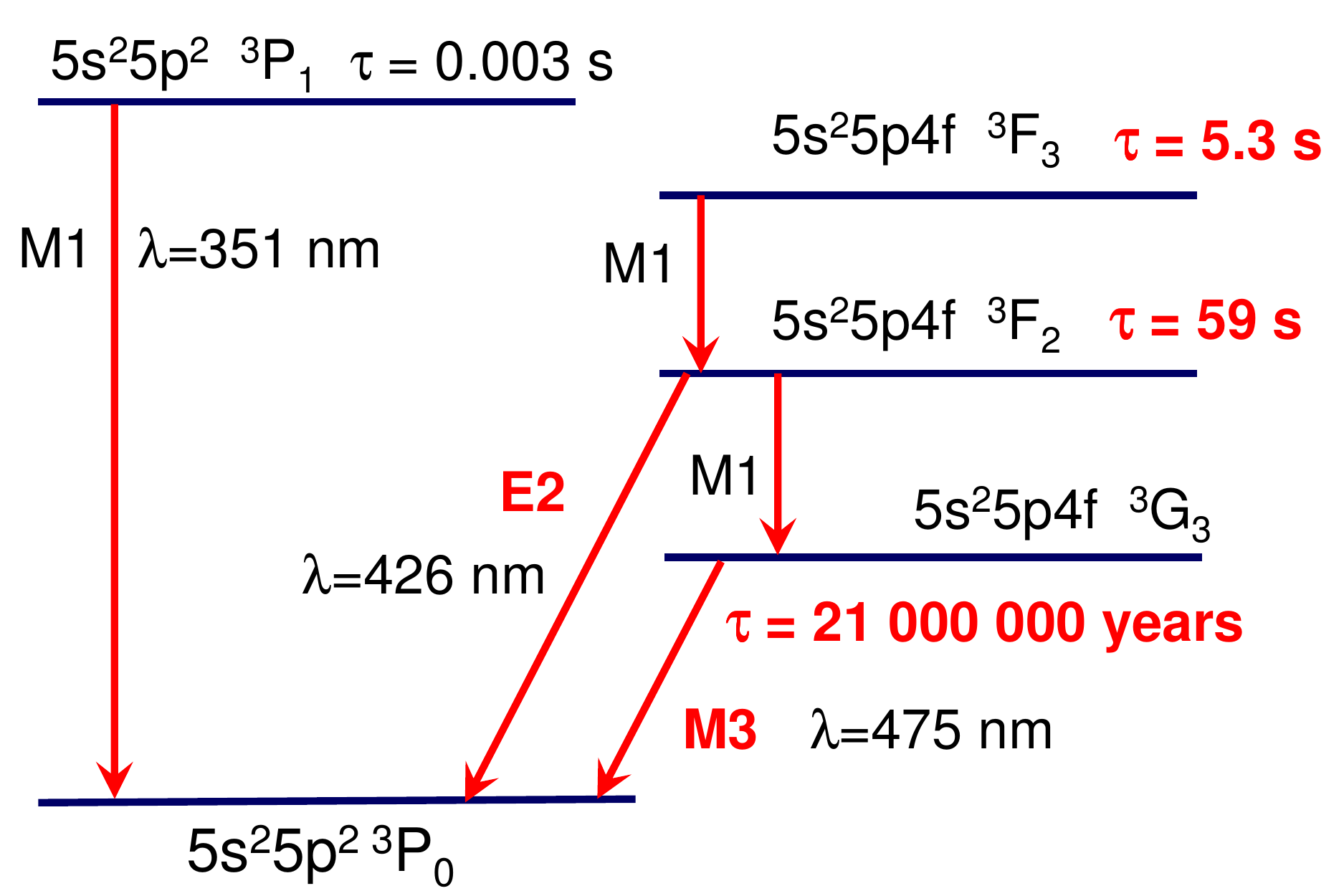}
  \caption{\normalsize{Energy levels and radiative lifetimes of low-lying levels of Sn-like Pr$^{9+}$.}
    \label{fig1}}
\end{figure}

Our study of the HCIs showed that some of these systems have several metastable states representing a level structure and other properties that are not present in any neutral atom and low-ionization state ions and may be advantageous not only for the development of atomic clocks but also for providing new possibilities for quantum information storage and processing. For example, we find that the first excited state in Sn-like Pr$^{9+}$  can be counted  as a second ``ground'' state with estimated lifetime of $10^{14}$ seconds since the fastest transition contributing to the lifetime is M3. The low-lying levels of Sn-like Pr$^{9+}$ ion and our estimates of  the radiative lifetimes are shown in Fig.~\ref{fig1} for illustration.  Much more compact size of the HCI electronic cloud may significantly reduce the decoherence processes due to undesired external perturbation, and extremely long-lived states may provide a resource for quantum memories. The large charge may allow for faster quantum logic gate operation~\cite{CM}.

While highly-charged ions lack strong electric-dipole transitions for laser cooling, sympathetic cooling may be employed similar to the case of Al$^+$ clock,  which is cooled using laser-cooled Be$^{+}$ or Mg$^+$ ion \cite{1}. The experimental investigations toward the sympathetic cooling of HCIs and the precision laser spectroscopy of forbidden transitions are currently in progress \cite{HobSolSuh11,AndCazNor13,SchVerWin12,VerSchWin13}. Recently, the evaporative cooling of Ar$^{16+}$ in a Penning trap was demonstrated~\cite{HobSolSuh11}. Storage and cooling of highly-charged ions require ultra-high vacuum levels that can be obtained by cryogenic methods. A linear Paul trap operating at 4~K capable of very long ion storage times of about 30~h was recently developed in ~\cite{SchVerWin12}.

{\bf Methods.} We use two different relativistic high-precision all-order approaches for all of the  calculations carried out in this work. The first approach is the relativistic linearized coupled-cluster  method that includes all single, double, and partial triple excitations of Dirac-Fock wave function~\cite{SafJoh08}. It has been extremely successful in predicting the properties of alkali-metal atoms and monovalent ions \cite{SafJoh08}. The second approach combines modified linearized coupled-cluster method with the configuration interaction (CI)  \cite{Koz04,SafKozJoh09}. This CI+all-order method yielded accurate atomic properties for a number of divalent systems and Tl having
three valence electrons
\cite{SafKozJoh09,SafKozCla11,PorSafKoz12,SafPorKoz12,SafKozCla12,SafKozSaf12,SafPorCla12}. This work presents  the first application of the CI+all-order method for 4-valence electron systems such as Sn-like ions. We demonstrated that the CI+all-order method can accurately predict properties of systems with four valence electrons (using test systems where the experimental results are known)  and designed efficient algorithms to construct 4-electron CI configuration spaces that contain all important sets of configurations.

An accurate calculation of the highly-charged ion energies and corresponding wavelengths not only requires the all-order methods to treat the Coulomb correlation effects,  but also an accurate treatment  of the Breit interaction, QED corrections, and inclusion of the high partial wave contribution. We treat the Breit interaction  on the same footing as the Coulomb interaction in the basis set.
The QED radiative corrections to the energy levels were included  using the method described in \cite{FlaGin05}.  The partial waves with $l_{max}=6$ are included in all summations in many-body perturbation theory or all-order terms, and $l>6$ contributions are extrapolated.  We find that inclusion of the higher partial waves with $l>6$ is very important for accurate description of the $4f$ states.  We illustrate the size of various contributions to the energies of Ag-like Nd$^{13+}$, In-like Pr$^{10+}$, and Sn-like Pr$^{9+}$ ions in Table~I of the supplementary material~\cite{EPAPS} to show that all of these contributions are essential for accurate determination of the positions of the energy levels and transition wavelengths of the HCIs. \begin{table} \caption{\label{tab1} Comparison of the energies of Ag-like  Nd$^{13+}$, Sm$^{15+}$, and In-like Ce$^{9+}$ ions relative to the ground state with experiment  \cite{ag-like-81,in-like-01}  and other theory ~\cite{DzuDerFla12}. Differences with experiment are given in cm$^{-1}$ and \% in columns ``Diff.''}
\begin{ruledtabular}
\begin{tabular}{llrrrrr}
\multicolumn{1}{c}{Ion}&
\multicolumn{1}{c}{Level}&
\multicolumn{1}{c}{Ref.~\cite{DzuDerFla12}}&
 \multicolumn{1}{c}{Expt.}&
\multicolumn{1}{c}{Present}&
\multicolumn{1}{c}{Diff.}&
\multicolumn{1}{c}{Diff.\%}\\
\hline   \\[-0.4pc]
Nd$^{13+}$ & $5s_{1/2}$&    0&        0&       0&   0 &         \\
           & $4f_{5/2}$&58897&    55870&  55706&  164& 0.29\% \\
           & $4f_{7/2}$&63613&    60300&  60134&  166& 0.28\% \\
           & $5p_{1/2}$&     &   185066&  185028&   38& 0.02\% \\
           & $5p_{3/2}$&     &   234864&  234887&  -23&-0.01\%  \\ [0.4pc]
Sm$^{15+}$ &$4f_{5/2}$ &    0&        0&     0  &      0  &          \\
           &$4f_{7/2}$ & 6806&     6555&  6444 &      111&   1.69\%   \\
           &$5s_{1/2}$ &55675&    60384& 60517&     -133&  -0.22\%    \\
           &$5p_{1/2}$ &     &   268488& 268604&     -116&  -0.04\%    \\
           &$5p_{3/2}$ &     &   333203& 333385&     -182&  -0.05\%     \\ [0.4pc]
Ce$^{9+}$  &$5p_{1/2}$&      &      0  &      0 &    0&          \\
           &$5p_{3/2}$&      & 33427   &  33450 & -23&  -0.07\%   \\
            &$4f_{5/2}$&      & 54947   &   54683&  264&   0.48\%  \\
           &$4f_{7/2}$&      & 57520   &   57235&  285&   0.50\%
\end{tabular}
\end{ruledtabular}
\end{table}

To further verify the accuracy of our calculations, we compared our values for the three ions where reliable spectra identification provides the experimental values of the energy levels. Comparison of the energies of Ag-like  Nd$^{13+}$, and Sm$^{15+}$ ions and In-like Ce$^{9+}$ relative to the ground state with experiment  \cite{ag-like-81}  and other theory ~\cite{DzuDerFla12} is given in Table~\ref{tab1}. Our values are in excellent agreement with experiment for all three cases.

{\bf Results.} In our study, we considered the following most essential properties for the experimental development: ground state ionization energies, energy levels, transition wavelengths, lifetimes, and sensitivity to the variation of the fine-structure constant $\alpha$. The sensitivity of the atomic transition frequency $\omega$  to the variation of $\alpha$ can be quantified using a coefficient $q$ defined as
$
 \omega(x)=\omega_0+qx,
$
  where
$
 x=\left(\frac{\alpha}{\alpha_0}\right)^2-1
$
and the frequency $\omega_0$ corresponds to the value of the fine-structure constant $\alpha_0$ at some initial point in time. In the experiment, the ratio of two frequencies, which is a dimensionless quantity, is studied over time. Therefore, it is preferable to select transitions with significantly different values of  $q$.  It is convenient to define dimensionless  enhancement factor $K=2q/\omega$ in order to compare the sensitivity to $\alpha$ between the transitions with significantly different frequencies as wide range of transition frequencies are valuable in HCIs. \begin{table} \caption{\label{tab2} Energies and $\alpha$-variation sensitivity coefficients $q$ for Ag-like and In-like ions relative to the ground state in cm$^{-1}$; $K=2q/\omega$ is the enhancement factor. Wavelengths $\lambda$ (in nm) for transitions to the ground states and total radiative lifetimes $\tau$ are listed. Experimental values are given for the wavelengths~\cite{ag-like-81,in-like-01} in Nd$^{13+}$, Sm$^{15+}$ and Ce$^{9+}$.}
\begin{ruledtabular}
\begin{tabular}{lrrrcc}
\multicolumn{1}{c}{Level}&
\multicolumn{1}{c}{Energy}&
\multicolumn{1}{c}{$q$}&
\multicolumn{1}{c}{$K$}&
 \multicolumn{1}{c}{$\lambda$}& \multicolumn{1}{c}{$\tau$}\\
\hline   \\[-0.4pc]
\multicolumn{2}{l}{\textbf{Ag-like Nd$^{13+}$}}  & \multicolumn{4}{l}{$5s_{1/2}$ ~ground state}   \\   [0.4pc]
 $4f_{5/2}$  &   55706   &    104229  &   3.7 &179.0  & $1.3\times10^6$~s \\
 $4f_{7/2}$  &   60134   &    108243  &   3.6 &165.8  & 0.996~s           \\
 $5p_{1/2}$  &   185028  &    15953   &   0.2 & 54.03 & 0.204~ns \\
 $5p_{3/2}$  &   234887  &    72079   &   0.6 & 42.58 & 0.098~ns \\ [0.4pc]
\multicolumn{2}{l}{\textbf{Ag-like Sm$^{15+}$}}  & \multicolumn{4}{l}{$4f_{5/2}$ ~ground state}   \\   [0.4pc]
  $4f_{7/2}$  &   6444    &    5910    &   1.8 & 1526  & 0.308~s  \\
  $5s_{1/2}$  &   60517   &    -134148 &   -4.4& 165.6 & $3.1\times10^5$~s \\
  $5p_{1/2}$  &   268604  &    -114999 &   -0.9& 37.25 & 0.167~ns \\
  $5p_{3/2}$  &   333385  &    -41477  &   -0.2& 30.01 & 0.0731~ns \\[0.4pc]
\multicolumn{2}{l}{\textbf{In-like Ce$^{9+}$}}  & \multicolumn{4}{l}{$5p_{1/2}$ ~ground state}   \\   [0.4pc]
     $5p_{3/2}$&  33450   &   37544   &   2.2&  299.2&0.0030~s    \\
     $4f_{5/2}$&  54683   &   62873   &   2.3&  182.0&0.0812~s    \\
     $4f_{7/2}$&  57235   &   65150   &   2.3&  173.9&2.18~s      \\[0.4pc]
 \multicolumn{2}{l}{\textbf{In-like Pr$^{10+}$}}  & \multicolumn{4}{l}{$5p_{1/2}$ ~ground state}   \\   [0.4pc]
      $4f_{5/2}$&  3702   &   73849   &   40  &2700(140)&8.5$\times10^{4}$~s  \\
      $4f_{7/2}$&  7031  &   76833   &   22  & 1422(40)&2.35~s \\
      $5p_{3/2}$&  39141      &   44098   &   2.3 & 255.5(3)&0.0018~s \\[0.4pc]
   \multicolumn{2}{l}{\textbf{In-like Nd$^{11+}$}}  & \multicolumn{4}{l}{$4f_{5/2}$ ~ground state}   \\   [0.4pc]
     $4f_{7/2}$&  4180    &   3785    &   1.8 &  2392(60)&1.19~s \\
     $5p_{1/2}$&  53684   &   -85692  &   -3.2& 186.3(1.7)&0.061~s   \\
     $5p_{3/2}$&  99066   &   -34349  &   -0.7&   100.9(5)& 0.00088~s
\end{tabular}
\end{ruledtabular}
\end{table}
\begin{table}
\caption{\label{tab4} Energies and $\alpha$-variation sensitivity coefficients $q$ for Cd-like, In-like, and Sn-like ions relative to the ground state in cm$^{-1}$; $K=2q/\omega$ is the enhancement factor. Wavelengths $\lambda$ (in nm) for transitions to the ground states and total radiative lifetimes $\tau$ are listed.}
\begin{ruledtabular}
\begin{tabular}{lrrrrc}
\multicolumn{1}{c}{Level}&
\multicolumn{1}{c}{Energy}&
\multicolumn{1}{c}{$q$}&
\multicolumn{1}{c}{$K$}&
 \multicolumn{1}{c}{$\lambda$}& \multicolumn{1}{c}{$\tau$}\\
\hline   \\[-0.4pc]
\multicolumn{2}{l}{\textbf{Cd-like Nd$^{12+}$}}  & \multicolumn{4}{l}{$5s^2 \ ^1S_0$ ~ground state}   \\   [0.4pc]
   $5s4f \ ^3F_2$  &   79469   &   101461  &   2.6 & 125.8(9)&8.5$\times10^{10}$~s  \\
       $5s4f \ ^3F_3$  &   80769   &   102325  &   2.5 & 123.8(9)&19.7~s       \\
       $5s4f \ ^3F_4$  &   83730   &   105340  &   2.5 & 119.4(9)&1.95~s       \\
       $5s5p \ ^3P_1$  &   168547  &   19465   &   0.2 & 59.3(3) &1.80~ns  \\[0.4pc]
 \multicolumn{2}{l}{  \textbf{Cd-like  Sm$^{14+}$}}  & \multicolumn{4}{l}{ $4f^2 \  ^3H_4$ ~ground state}   \\   [0.4pc]
        $5s4f \ ^3F_2$  &   2172  &   -127720 &   -118&4600      &5.6$\times10^{13}$~s    \\
        $5s4f \ ^3F_3$  &   3826  &   -126746 &   -66 &2614(470) &8.515~s        \\
        $4f^2 \ ^3H_5$  &   4939  &   4917    &   2.0 &2025(40)  &0.313~s        \\
        $5s4f \ ^3F_4$  &   8463  &   -121952 &   -29 &1182(100) &0.556~s\\[0.4pc]
   \multicolumn{2}{l}{\textbf{In-like  Sm$^{13+}$}}  & \multicolumn{4}{l}{$5s^2 4f_{5/2}~^2F_{5/2}$ ~ground state}   \\   [0.4pc]
               $5s^2 4f  ~^2F_{7/2}$ &      6203    &   5654    &   1.8& 1612(28)&0.367~s \\
               $4f^2 5s  ~^4H_{7/2}$ &      20254   &   123621  &   12 &  494(22)&0.133~s \\
               $4f^2 5s  ~^4H_{9/2}$ &      22519   &   125397  &   11 &  444(18)&0.141~s \\
               $4f^25s   ~^4H_{11/2}$&      25904   &   128875  &   10 &  386(14)&0.576~s\\ [0.4pc]
   \multicolumn{2}{l}{\textbf{Sn-like Pr$^{9+}$}}  & \multicolumn{4}{l}{$5p^2$~$^3P_0$ ~ground state}   \\   [0.4pc]
     $5p4f~^3G_3$     &   20216   &   42721   &   4.2& 475(18)& 6.6$\times10^{14}$~s  \\
     $5p4f~^3F_2$     &   22772   &   42865   &   3.8& 426(13)& 59.0~s                \\
     $5p4f~^3F_3$     &   25362   &   47076   &   3.7& 382(12)& 5.33~s                \\
     $5p4f~^3F_4$     &   27536   &   37197   &   2.7& 352(11)& 0.234~s               \\
     $5p^2$~$^3P_1$   &   28512   &   47483   &   3.3& 351.2(5)&0.00296~s                    \\[0.4pc]
 \multicolumn{2}{l}{\textbf{Sn-like Nd$^{10+}$}}  & \multicolumn{4}{l}{$4f^2~J=4$ ~ground state}   \\   [0.4pc]
    $5p4f~\textrm{J=3}$  &   1564    &   -81052  &   -104 &         &  1.6$\times10^{4}$~s   \\
    $4f^2~\textrm{J=5}$  &   3059    &   3113    &   2.0  &         & 1.4~s  \\
    $5p4f~\textrm{J=2}$  &   5060    &   -60350  &   -24  &2200(430)&  25~s\\
    $4f^2~\textrm{J=6}$  &   6222    &   5930    &   1.9  &1610(110)& 1.4~s \\
    $5p4f~\textrm{J=3}$  &   7095    &   -63285  &   -18  &1480(240)& 3.9~s
\end{tabular}
\end{ruledtabular}
\end{table}
\begin{table}
\caption{\label{tabIP} Ionization energies (IE) of Ag-like and In-like ions in cm$^{-1}$ and eV. Comparison with experiment \cite{nist-web}
is given where available. Ground state configurations of the valence electron are listed.}
\begin{ruledtabular}
\begin{tabular}{llcrrr}
\multicolumn{2}{c}{Ion}&\multicolumn{1}{c}{Level}& \multicolumn{1}{c}{IE(cm$^{-1}$)}&
\multicolumn{1}{c}{IE(eV)}& \multicolumn{1}{c}{Expt. (eV)} \\  \\[-0.6pc] \hline
Ag-like& Ba$^{9+} $&$5s $& 1181600 &146.5 &  146.52(12) \\
Ag-like& Nd$^{13+}$&$5s $& 1960539 &243.1 &             \\
Ag-like& Sm$^{15+}$&$4f_{5/2} $& 2471848 &306.5 &             \\
In-like& Ca$^{6+} $&$5p_{1/2} $& 671692  &83.3  &  82.9(2.0)  \\
In-like& Ba$^{7+} $&$5p_{1/2} $& 816588  &101.2 &  101.0(2.1) \\
In-like& Ce$^{9+} $&$5p_{1/2} $& 1138158 &141.1 &            \\
In-like& Pr$^{10+}$&$5p_{1/2} $& 1314931 &163.0 &           \\
In-like& Nd$^{11+}$&$4f_{5/2} $& 1554148 &192.7 &           \\
\end{tabular}
\end{ruledtabular}
\end{table}
To calculate the value of the sensitivity coefficient $q$, we carry out three calculations with different values of $\alpha$ for every ion considered in this work.  In the first calculation, current CODATA value of $\alpha$ ~\cite{MohTayNew11} is used. In the other two calculations, the value of $\alpha^2$ is varied by $\pm$1\%. The value of $q$ is then determined as a numerical derivative.

The results for Ag-like ions and In-like ions that can be considered ``monovalent'', i.e. with one valence electron above the closed $5s^2$ shell are given in Table~\ref{tab2}.  Energies and $\alpha$-variation sensitivity coefficients $q$ relative to the ground state in cm$^{-1}$, enhancement factor $K=2q/\omega$,  wavelengths (in nm) for transitions to the ground states and total radiative lifetimes $\tau$ are listed. Experimental values are given for the wavelengths~\cite{ag-like-81,in-like-01} of the first three ions. We developed several methods to evaluate the uncertainties  in the values of the theoretical wavelengths based on the size of all four corrections described above and comparison with experiment for previous ions of the corresponding isoelectronic sequences, for example, Ba$^{6+}$ -- Ba$^{9+}$. The uncertainties for the wavelengths transitions to the ground states are listed in parenthesis for In-like ions.

The $5s-4f$ level crossing in Ag-like isoelectronic sequence happens from Nd$^{13+}$ to Sm$^{15+}$.  The Pm$^{14+}$ has no stable isotopes and  the $5s$ and $4f_{5/2}$ states in Pm$^{14+}$ are separated by only about 300~cm$^{-1}$. Nd$^{13+}$ represents a particularly attractive case since   the strongest transition from the metastable $4f_{5/2}$ level of this ion is E3, resulting in the extremely long lifetime of more than 15~days. Therefore, this system may be considered to have two ground states.   While Nd$^{13+}$ transition $5s-4f$ wavelength is in UV, another ion can be used to probe the transitions as in Al$^+$ clock scheme which also has a clock transition in UV. Many of low-lying transitions in all ions considered here have large $q$  and $K$  coefficients indicating large sensitivities to $\alpha$-variation. Our  calculations of $K$ for infrared transitions in Pr$^{10+}$ give  $K=22$ and $K=40$. For comparison, $K=3$ \cite{DzuFla08} for UV transition in Hg$^{+}$ which provided the most precise current test of $\alpha$-variation in Al$^+$/Hg$^+$ clocks carried out in ~\cite{RosHumSch08}. The Al$^+$ ion is light and does not produce additional enhancement of sensitivity to $\alpha$-variation.

 There are two level crossings of interest for the present work in In-like isoelectronic sequence, $5p-4f$ and $4f-5s$.  The first one happens for Pr$^{10+}$ and Nd$^{11+}$ and leads to change of level order from   $5p$, $4f$ to $4f$, $5p$.  Pr$^{10+}$ represents a particularly attractive case where both $4f_j$ levels are located between the $5p_{1/2}$ and $5p_{3/2}$ fine structure multiplet, making $4f_{5/2}$ a very long-lived metastable level with 1~day radiative lifetime.  The order of levels changes again for Sm$^{13+}$, where $5s4f^2$ configuration becomes the closest to the ground $5s^24f_{j}$  fine-structure multiplet. This leads to a very interesting level structure with a metastable $5s4f^2$ $J=7/2$ level in the optical transition range to both ground and excited  $5s^24f_{7/2}$ levels of the fine-structure multiplet.  The results for this In-like ion are listed in Table~\ref{tab4} together with the results of other multi-valence electron HCIs in Cd-like and Sn-like isoelectronic sequences.

The $5s-4f$ level crossing in Cd-like ions happens for Nd$^{12+}$ - Sm$^{14+}$ ions.  The strongest transition from the first excited levels of both ions is M2,   resulting in the extremely long lifetimes. The case of Nd$^{12+}$ is very similar to Ag-like Nd$^{13+}$ but the wavelengths are further in UV. We note that $q$ values are positive for Nd$^{12+}$  and negative for Sm$^{14+}$  which creates additional enhancements for $\alpha$-variation search if the relative transition frequencies in Nd$^{12+}$/Sm$^{14+}$ are monitored.  Similar cases are found in other isoelectronic sequences studied here because the energy levels change order    with the level crossings and  the values of $q$ in ions before and after the crossing  have an opposite sign.

The ions of interest in the Sn-like isoelectronic sequence are Pr$^{9+}$ and Nd$^{10+}$. The case of Pr$^{9+}$ is particularly interesting, since the lowest metastable state, $5p4f$~$J=3$ is extremely long-lived with transition to the ground state being in the optical range, 495~nm (see Fig.~\ref{fig1}). The strongest allowed transition is M3 making this ion a unique system. Next two levels also have optical transitions to the ground state and are metastable  with 59~s and 5.3~s lifetimes.
Moreover, there is a relatively strong M1 transition to the ground
state from $5p^2$~$^3P_1$  at 351~nm that may be potentially used for
cooling and probing. The ground and first excited states of Nd$^{10+}$
are extremely close and the resulting uncertainty is on the order of
the transition energy.
While our calculations place $4f^2$ to be the ground state, the higher-order corrections are particularly large in this case, almost 3 times that of the transition energy, which might lead to the placement of the $5p4f$ $J=3$ as the ground state.

The ionization potentials for Ag-like and In-like ions are given in Table~\ref{tabIP}. We note that their values  mainly depends on the degree of ionization (147~eV for Ag-like Ba$^{9+}$ and 141~eV for In-like Ce$^{9+}$) so this table may be used to estimate the ionization potentials for Cd-like and Sn-like ions as well.

In summary, we provided a clear direction for the further experimental study of highly-charged ions for the development of next-generation atomic clocks and search for $\alpha$-variation. We conducted a first high-precision study of such systems  as well as demonstrated the first treatment  of any systems with four valence electrons (Sn-like ions) using the all-order methodology. We identified  \textit{all} ions that fit the criteria for the most straightforward experimental exploration including sensitivity to $\alpha$-variation. We also proposed the use of  highly-charged ions  for quantum information research. The ions listed in this work present a completely unexplored resource for quantum information due to their unique atomic properties and potential reduced sensitivity to decoherence effects. We hope that this work will lay the foundation for the future development of this field.

We thank C. W. Clark, C. Monroe, J. Tan, Yu. Ralchenko, and P. Beiersdorfer for useful discussions. This work was supported in part by US NSF Grant No.\ PHY-1212442. M.G.K. acknowledges support from RFBR Grant No.\ 14-02-00241.
M.S.S. thanks School of Physics at UNSW, Sydney, Australia for hospitality and acknowledges support from Gordon Godfrey Fellowship, UNSW.
The work is partly supported by the Australian Research Council.

\newpage
\begin{center}
\large{\textbf{Supplemental Material}}
\end{center}

Accurate calculation of the highly-charged ion properties not only require the all-order methods to treat the Coulomb correlation effects,  but also accurate treatment  of the Breit interaction, QED corrections, and inclusion of the high partial wave contribution.  We illustrate the size of various contributions to the energies of Ag-like Nd$^{13+}$, In-like Pr$^{10+}$, and Sn-like Pr$^{9+}$ ions in Table~\ref{tab1}. The contribution of the QED correction for the ions of interest to this work is only significant for the configurations that contain valence  $5s$ state. Therefore, the QED can be omitted for ions where none of the low-lying configurations  contain $5s$ valence state.  All energy values are counted from the corresponding ground state energies.  Contributions to the energies of highly-charged ions (in cm$^{-1}$) from higher-order Coulomb correlation (above  second-order MBPT or CI+MBPT), estimated contributions of  higher partial waves ($l>6$), Breit interaction, and QED are given separately in columns HO, $l>6 $, Breit, and QED, respectively. Results of 4-valence electron calculation are given for Pr$^{9+}$. CI+MBPT results are given in column MBPT for Pr ions.

\begin{table} [b]
\caption{\label{tab1}  Contributions to the energies of highly-charged ions (in cm$^{-1}$)
from higher-order Coulomb correlation (above second-order MBPT or CI+MBPT),
estimated contributions of  higher partial waves ($l>6$),
Breit interaction, and QED are given separately in columns HO, $l>6
$, Breit, and QED, respectively.
Results of 4-valence electron calculation are given for Pr$^{9+}$. CI+MBPT results are given
in column MBPT for Pr ions.}
\begin{ruledtabular}
\begin{tabular}{llcrrrrr}
\multicolumn{1}{l}{Ion}&
\multicolumn{1}{l}{Level}&
 \multicolumn{1}{c}{MBPT}&
\multicolumn{1}{c}{HO}&
\multicolumn{1}{c}{$l>6$}&
\multicolumn{1}{c}{Breit}&
\multicolumn{1}{c}{QED}&
\multicolumn{1}{c}{Final}\\
\hline   \\[-0.4pc]
Nd$^{13+}$ &$5s_{1/2}$&      0&   0   &     0 &      0&   0     &       0 \\
           &$4f_{5/2}$&  58596& 761   & -1247 &  -1421&   -983  &   55706  \\
           &$4f_{7/2}$&  63429& 671   & -1238 &  -1767&  -961   &   60134  \\
           &$5p_{1/2}$& 185876&-492   &   -33 &    560&   -883  &  185028 \\
           &$5p_{3/2}$& 236463&-747 &   -14 &    -14&   -801  &  234887 \\ [0.5pc]
Pr$^{10+}$ &$5p_{1/2}$&      0 &     0  &      0&      0&&  0   \\
           &$4f_{5/2}$&  3471  & 2821   &  -1063&  -1797&& 3702\\
           &$4f_{7/2}$&  7136  & 2761   &  -1057&  -2079&& 7031\\
           &$5p_{3/2}$&  39745 & -154   &   14  &   -464&& 39141\\  [0.5pc]
Pr$^{9+}$  &$5s^25p^2$~$^3P_0$&    0 &0     &     0&     0 &&  0      \\
           &$5s^25p4f~^3G_3$  &21865 & 2810 & -1032& -1748 &&  21895  \\
           &$5s^25p4f~^3F_2$  &24172 & 2291 & -829 & -1435 &&  24199   \\
           &$5s^25p4f~^3F_3$  &27233 & 2804 & -1026& -2009 &&  27002
\end{tabular}
\end{ruledtabular}
\end{table}

Table~\ref{tab1} illustrates that all of these contributions are essential for accurate determination of the positions of the energy levels and
transition wavelength of highly-charged ions.


\begin{thebibliography}{39}
\expandafter\ifx\csname natexlab\endcsname\relax\def\natexlab#1{#1}\fi
\expandafter\ifx\csname bibnamefont\endcsname\relax
  \def\bibnamefont#1{#1}\fi
\expandafter\ifx\csname bibfnamefont\endcsname\relax
  \def\bibfnamefont#1{#1}\fi
\expandafter\ifx\csname citenamefont\endcsname\relax
  \def\citenamefont#1{#1}\fi
\expandafter\ifx\csname url\endcsname\relax
  \def\url#1{\texttt{#1}}\fi
\expandafter\ifx\csname urlprefix\endcsname\relax\def\urlprefix{URL }\fi
\providecommand{\bibinfo}[2]{#2}
\providecommand{\eprint}[2][]{\url{#2}}

\bibitem[{1()}]{1}
\bibinfo{note}{C. W. Chou {\it et al.}, Phys. Rev. Lett. {\bf 104}, 070802
  (2010).}

\bibitem[{2()}]{2}
\bibinfo{note}{N. Hinkley {\it et al.}, Science {\bf 341}, 1215 (2013).}

\bibitem[{3()}]{3}
\bibinfo{note}{B. J. Bloom {\it et al.}, Nature, {\bf 506}, 71 (2014).}

\bibitem[{\citenamefont{{Lemke} et~al.}(2009)}]{Ybclock}
\bibinfo{author}{\bibfnamefont{N.~D.} \bibnamefont{{Lemke}}}
  \bibnamefont{et~al.}, \bibinfo{journal}{Phys. Rev. Lett.}
  \textbf{\bibinfo{volume}{103}}, \bibinfo{eid}{063001} (\bibinfo{year}{2009}).

\bibitem[{\citenamefont{{Madej} et~al.}(2012)\citenamefont{{Madej}, {Dub{\'e}},
  {Zhou}, {Bernard}, and {Gertsvolf}}}]{MadDubZho12}
\bibinfo{author}{\bibfnamefont{A.~A.} \bibnamefont{{Madej}}},
  \bibinfo{author}{\bibfnamefont{P.}~\bibnamefont{{Dub{\'e}}}},
  \bibinfo{author}{\bibfnamefont{Z.}~\bibnamefont{{Zhou}}},
  \bibinfo{author}{\bibfnamefont{J.~E.} \bibnamefont{{Bernard}}},
  \bibnamefont{and}
  \bibinfo{author}{\bibfnamefont{M.}~\bibnamefont{{Gertsvolf}}},
  \bibinfo{journal}{Phys. Rev. Lett.} \textbf{\bibinfo{volume}{109}},
  \bibinfo{eid}{203002} (\bibinfo{year}{2012}).

\bibitem[{\citenamefont{{Uzan}}(2003)}]{Uza03}
\bibinfo{author}{\bibfnamefont{J.-P.} \bibnamefont{{Uzan}}},
  \bibinfo{journal}{Rev. Mod. Phys.} \textbf{\bibinfo{volume}{75}},
  \bibinfo{pages}{403} (\bibinfo{year}{2003}).

\bibitem[{\citenamefont{{Murphy} et~al.}(2003)\citenamefont{{Murphy}, {Webb},
  and {Flambaum}}}]{MurWebFla03}
\bibinfo{author}{\bibfnamefont{M.~T.} \bibnamefont{{Murphy}}},
  \bibinfo{author}{\bibfnamefont{J.~K.} \bibnamefont{{Webb}}},
  \bibnamefont{and} \bibinfo{author}{\bibfnamefont{V.~V.}
  \bibnamefont{{Flambaum}}}, \bibinfo{journal}{Mon. Not. R. Astron. Soc.}
  \textbf{\bibinfo{volume}{345}}, \bibinfo{pages}{609} (\bibinfo{year}{2003}).

\bibitem[{\citenamefont{{Webb} et~al.}(2011)\citenamefont{{Webb}, {King},
  {Murphy}, {Flambaum}, {Carswell}, and {Bainbridge}}}]{WebKinMur11}
\bibinfo{author}{\bibfnamefont{J.~K.} \bibnamefont{{Webb}}},
  \bibinfo{author}{\bibfnamefont{J.~A.} \bibnamefont{{King}}},
  \bibinfo{author}{\bibfnamefont{M.~T.} \bibnamefont{{Murphy}}},
  \bibinfo{author}{\bibfnamefont{V.~V.} \bibnamefont{{Flambaum}}},
  \bibinfo{author}{\bibfnamefont{R.~F.} \bibnamefont{{Carswell}}},
  \bibnamefont{and} \bibinfo{author}{\bibfnamefont{M.~B.}
  \bibnamefont{{Bainbridge}}}, \bibinfo{journal}{Phys. Rev. Lett.}
  \textbf{\bibinfo{volume}{107}}, \bibinfo{eid}{191101} (\bibinfo{year}{2011}).

\bibitem[{\citenamefont{{Rahmani} et~al.}(2012)\citenamefont{{Rahmani},
  {Srianand}, {Gupta}, {Petitjean}, {Noterdaeme}, and {V{\'a}squez}}}]{RSGP12}
\bibinfo{author}{\bibfnamefont{H.}~\bibnamefont{{Rahmani}}},
  \bibinfo{author}{\bibfnamefont{R.}~\bibnamefont{{Srianand}}},
  \bibinfo{author}{\bibfnamefont{N.}~\bibnamefont{{Gupta}}},
  \bibinfo{author}{\bibfnamefont{P.}~\bibnamefont{{Petitjean}}},
  \bibinfo{author}{\bibfnamefont{P.}~\bibnamefont{{Noterdaeme}}},
  \bibnamefont{and} \bibinfo{author}{\bibfnamefont{D.~A.}
  \bibnamefont{{V{\'a}squez}}}, \bibinfo{journal}{Mon. Not. R. Astron. Soc.}
  \textbf{\bibinfo{volume}{425}}, \bibinfo{pages}{556} (\bibinfo{year}{2012}).

\bibitem[{\citenamefont{{Berengut} and {Flambaum}}(2011)}]{BerFla11}
\bibinfo{author}{\bibfnamefont{J.~C.} \bibnamefont{{Berengut}}}
  \bibnamefont{and} \bibinfo{author}{\bibfnamefont{V.~V.}
  \bibnamefont{{Flambaum}}}, \bibinfo{journal}{J. Phys. Conf. Ser.}
  \textbf{\bibinfo{volume}{264}}, \bibinfo{pages}{012010}
  (\bibinfo{year}{2011}).

\bibitem[{\citenamefont{Berengut and Flambaum}(2012)}]{BerFla2012}
\bibinfo{author}{\bibfnamefont{J.~C.} \bibnamefont{Berengut}} \bibnamefont{and}
  \bibinfo{author}{\bibfnamefont{V.~V.} \bibnamefont{Flambaum}},
  \bibinfo{journal}{EPL} \textbf{\bibinfo{volume}{97}}, \bibinfo{pages}{20006}
  (\bibinfo{year}{2012}).

\bibitem[{\citenamefont{{Rosenband} et~al.}(2008)}]{RosHumSch08}
\bibinfo{author}{\bibfnamefont{T.}~\bibnamefont{{Rosenband}}}
  \bibnamefont{et~al.}, \bibinfo{journal}{Science}
  \textbf{\bibinfo{volume}{319}}, \bibinfo{pages}{1808} (\bibinfo{year}{2008}).

\bibitem[{\citenamefont{Dzuba et~al.}(2012)\citenamefont{Dzuba, Derevianko, and
  Flambaum}}]{DDf12}
\bibinfo{author}{\bibfnamefont{V.~A.} \bibnamefont{Dzuba}},
  \bibinfo{author}{\bibfnamefont{A.}~\bibnamefont{Derevianko}},
  \bibnamefont{and} \bibinfo{author}{\bibfnamefont{V.~V.}
  \bibnamefont{Flambaum}}, \bibinfo{journal}{Phys. Rev. A}
  \textbf{\bibinfo{volume}{86}}, \bibinfo{pages}{054501}
  (\bibinfo{year}{2012}).

\bibitem[{\citenamefont{{Derevianko} et~al.}(2012)\citenamefont{{Derevianko},
  {Dzuba}, and {Flambaum}}}]{DerDzuFla12}
\bibinfo{author}{\bibfnamefont{A.}~\bibnamefont{{Derevianko}}},
  \bibinfo{author}{\bibfnamefont{V.~A.} \bibnamefont{{Dzuba}}},
  \bibnamefont{and} \bibinfo{author}{\bibfnamefont{V.~V.}
  \bibnamefont{{Flambaum}}}, \bibinfo{journal}{Phys. Rev. Lett.}
  \textbf{\bibinfo{volume}{109}}, \bibinfo{eid}{180801} (\bibinfo{year}{2012}).

\bibitem[{\citenamefont{{Berengut} et~al.}(2010)\citenamefont{{Berengut},
  {Dzuba}, and {Flambaum}}}]{BerDzuFla10}
\bibinfo{author}{\bibfnamefont{J.~C.} \bibnamefont{{Berengut}}},
  \bibinfo{author}{\bibfnamefont{V.~A.} \bibnamefont{{Dzuba}}},
  \bibnamefont{and} \bibinfo{author}{\bibfnamefont{V.~V.}
  \bibnamefont{{Flambaum}}}, \bibinfo{journal}{Phys. Rev. Lett.}
  \textbf{\bibinfo{volume}{105}}, \bibinfo{eid}{120801} (\bibinfo{year}{2010}).

\bibitem[{\citenamefont{{Berengut} et~al.}(2011)\citenamefont{{Berengut},
  {Dzuba}, {Flambaum}, and {Ong}}}]{BerDzuFla11b}
\bibinfo{author}{\bibfnamefont{J.~C.} \bibnamefont{{Berengut}}},
  \bibinfo{author}{\bibfnamefont{V.~A.} \bibnamefont{{Dzuba}}},
  \bibinfo{author}{\bibfnamefont{V.~V.} \bibnamefont{{Flambaum}}},
  \bibnamefont{and} \bibinfo{author}{\bibfnamefont{A.}~\bibnamefont{{Ong}}},
  \bibinfo{journal}{Phys. Rev. Lett.} \textbf{\bibinfo{volume}{106}},
  \bibinfo{eid}{210802} (\bibinfo{year}{2011}).

\bibitem[{\citenamefont{{Dzuba} et~al.}(2012)\citenamefont{{Dzuba},
  {Derevianko}, and {Flambaum}}}]{DzuDerFla12}
\bibinfo{author}{\bibfnamefont{V.~A.} \bibnamefont{{Dzuba}}},
  \bibinfo{author}{\bibfnamefont{A.}~\bibnamefont{{Derevianko}}},
  \bibnamefont{and} \bibinfo{author}{\bibfnamefont{V.~V.}
  \bibnamefont{{Flambaum}}}, \bibinfo{journal}{\pra}
  \textbf{\bibinfo{volume}{86}}, \bibinfo{eid}{054502} (\bibinfo{year}{2012}).

\bibitem[{\citenamefont{{Berengut} et~al.}(2012)\citenamefont{{Berengut},
  {Dzuba}, {Flambaum}, and {Ong}}}]{BerDzuFla12}
\bibinfo{author}{\bibfnamefont{J.~C.} \bibnamefont{{Berengut}}},
  \bibinfo{author}{\bibfnamefont{V.~A.} \bibnamefont{{Dzuba}}},
  \bibinfo{author}{\bibfnamefont{V.~V.} \bibnamefont{{Flambaum}}},
  \bibnamefont{and} \bibinfo{author}{\bibfnamefont{A.}~\bibnamefont{{Ong}}},
  \bibinfo{journal}{\pra} \textbf{\bibinfo{volume}{86}}, \bibinfo{eid}{022517}
  (\bibinfo{year}{2012}), \eprint{1206.0534}.

\bibitem[{CM()}]{CM}
\bibinfo{note}{C. Monroe, private communication}.

\bibitem[{\citenamefont{{Hobein} et~al.}(2011)\citenamefont{{Hobein},
  {Solders}, {Suhonen}, {Liu}, and {Schuch}}}]{HobSolSuh11}
\bibinfo{author}{\bibfnamefont{M.}~\bibnamefont{{Hobein}}},
  \bibinfo{author}{\bibfnamefont{A.}~\bibnamefont{{Solders}}},
  \bibinfo{author}{\bibfnamefont{M.}~\bibnamefont{{Suhonen}}},
  \bibinfo{author}{\bibfnamefont{Y.}~\bibnamefont{{Liu}}}, \bibnamefont{and}
  \bibinfo{author}{\bibfnamefont{R.}~\bibnamefont{{Schuch}}},
  \bibinfo{journal}{Phys. Rev. Lett.} \textbf{\bibinfo{volume}{106}},
  \bibinfo{eid}{013002} (\bibinfo{year}{2011}).

\bibitem[{\citenamefont{{Andelkovic} et~al.}(2013)}]{AndCazNor13}
\bibinfo{author}{\bibfnamefont{Z.}~\bibnamefont{{Andelkovic}}}
  \bibnamefont{et~al.}, \bibinfo{journal}{Phys. Rev. A}
  \textbf{\bibinfo{volume}{87}}, \bibinfo{eid}{033423} (\bibinfo{year}{2013}).

\bibitem[{\citenamefont{{Schwarz} et~al.}(2012)}]{SchVerWin12}
\bibinfo{author}{\bibfnamefont{M.}~\bibnamefont{{Schwarz}}}
  \bibnamefont{et~al.}, \bibinfo{journal}{Rev. Sci. Instrum.}
  \textbf{\bibinfo{volume}{83}}, \bibinfo{pages}{083115}
  (\bibinfo{year}{2012}).

\bibitem[{\citenamefont{{Versolato} et~al.}(2013)}]{VerSchWin13}
\bibinfo{author}{\bibfnamefont{O.~O.} \bibnamefont{{Versolato}}}
  \bibnamefont{et~al.}, \bibinfo{journal}{Hyperfine Interact.}
  \textbf{\bibinfo{volume}{214}}, \bibinfo{pages}{189} (\bibinfo{year}{2013}).

\bibitem[{\citenamefont{Safronova and Johnson}(2008)}]{SafJoh08}
\bibinfo{author}{\bibfnamefont{M.~S.} \bibnamefont{Safronova}}
  \bibnamefont{and} \bibinfo{author}{\bibfnamefont{W.~R.}
  \bibnamefont{Johnson}}, \bibinfo{journal}{Adv. At. Mol. Opt. Phys.}
  \textbf{\bibinfo{volume}{55}}, \bibinfo{pages}{191} (\bibinfo{year}{2008}).

\bibitem[{\citenamefont{Kozlov}(2004)}]{Koz04}
\bibinfo{author}{\bibfnamefont{M.~G.} \bibnamefont{Kozlov}},
  \bibinfo{journal}{Int. J. Quant. Chem.} \textbf{\bibinfo{volume}{100}},
  \bibinfo{pages}{336} (\bibinfo{year}{2004}).

\bibitem[{\citenamefont{{Safronova} et~al.}(2009)\citenamefont{{Safronova},
  {Kozlov}, {Johnson}, and {Jiang}}}]{SafKozJoh09}
\bibinfo{author}{\bibfnamefont{M.~S.} \bibnamefont{{Safronova}}},
  \bibinfo{author}{\bibfnamefont{M.~G.} \bibnamefont{{Kozlov}}},
  \bibinfo{author}{\bibfnamefont{W.~R.} \bibnamefont{{Johnson}}},
  \bibnamefont{and} \bibinfo{author}{\bibfnamefont{D.}~\bibnamefont{{Jiang}}},
  \bibinfo{journal}{Phys. Rev. A} \textbf{\bibinfo{volume}{80}},
  \bibinfo{eid}{012516} (\bibinfo{year}{2009}).

\bibitem[{\citenamefont{{Safronova} et~al.}(2011)\citenamefont{{Safronova},
  {Kozlov}, and {Clark}}}]{SafKozCla11}
\bibinfo{author}{\bibfnamefont{M.~S.} \bibnamefont{{Safronova}}},
  \bibinfo{author}{\bibfnamefont{M.~G.} \bibnamefont{{Kozlov}}},
  \bibnamefont{and} \bibinfo{author}{\bibfnamefont{C.~W.}
  \bibnamefont{{Clark}}}, \bibinfo{journal}{Phys. Rev. Lett.}
  \textbf{\bibinfo{volume}{107}}, \bibinfo{eid}{143006} (\bibinfo{year}{2011}).

\bibitem[{\citenamefont{{Porsev} et~al.}(2012)\citenamefont{{Porsev},
  {Safronova}, and {Kozlov}}}]{PorSafKoz12}
\bibinfo{author}{\bibfnamefont{S.~G.} \bibnamefont{{Porsev}}},
  \bibinfo{author}{\bibfnamefont{M.~S.} \bibnamefont{{Safronova}}},
  \bibnamefont{and} \bibinfo{author}{\bibfnamefont{M.~G.}
  \bibnamefont{{Kozlov}}}, \bibinfo{journal}{Phys. Rev. Lett.}
  \textbf{\bibinfo{volume}{108}}, \bibinfo{eid}{173001} (\bibinfo{year}{2012}).

\bibitem[{\citenamefont{{Safronova} et~al.}(2012)\citenamefont{{Safronova},
  {Porsev}, {Kozlov}, and {Clark}}}]{SafPorKoz12}
\bibinfo{author}{\bibfnamefont{M.~S.} \bibnamefont{{Safronova}}},
  \bibinfo{author}{\bibfnamefont{S.~G.} \bibnamefont{{Porsev}}},
  \bibinfo{author}{\bibfnamefont{M.~G.} \bibnamefont{{Kozlov}}},
  \bibnamefont{and} \bibinfo{author}{\bibfnamefont{C.~W.}
  \bibnamefont{{Clark}}}, \bibinfo{journal}{Phys. Rev. A}
  \textbf{\bibinfo{volume}{85}}, \bibinfo{eid}{052506} (\bibinfo{year}{2012}).

\bibitem[{\citenamefont{Safronova et~al.}(2012)\citenamefont{Safronova, Kozlov,
  and Clark}}]{SafKozCla12}
\bibinfo{author}{\bibfnamefont{M.~S.} \bibnamefont{Safronova}},
  \bibinfo{author}{\bibfnamefont{M.~G.} \bibnamefont{Kozlov}},
  \bibnamefont{and} \bibinfo{author}{\bibfnamefont{C.~W.} \bibnamefont{Clark}},
  \bibinfo{journal}{IEEE T. Ultrason. Ferr.} \textbf{\bibinfo{volume}{59}},
  \bibinfo{pages}{439} (\bibinfo{year}{2012}).

\bibitem[{\citenamefont{{Safronova}
  et~al.}(2012{\natexlab{a}})\citenamefont{{Safronova}, {Kozlov}, and
  {Safronova}}}]{SafKozSaf12}
\bibinfo{author}{\bibfnamefont{M.~S.} \bibnamefont{{Safronova}}},
  \bibinfo{author}{\bibfnamefont{M.~G.} \bibnamefont{{Kozlov}}},
  \bibnamefont{and} \bibinfo{author}{\bibfnamefont{U.~I.}
  \bibnamefont{{Safronova}}}, \bibinfo{journal}{Phys. Rev. A}
  \textbf{\bibinfo{volume}{85}}, \bibinfo{eid}{012507}
  (\bibinfo{year}{2012}{\natexlab{a}}).

\bibitem[{\citenamefont{{Safronova}
  et~al.}(2012{\natexlab{b}})\citenamefont{{Safronova}, {Porsev}, and
  {Clark}}}]{SafPorCla12}
\bibinfo{author}{\bibfnamefont{M.~S.} \bibnamefont{{Safronova}}},
  \bibinfo{author}{\bibfnamefont{S.~G.} \bibnamefont{{Porsev}}},
  \bibnamefont{and} \bibinfo{author}{\bibfnamefont{C.~W.}
  \bibnamefont{{Clark}}}, \bibinfo{journal}{Phys. Rev. Lett.}
  \textbf{\bibinfo{volume}{109}}, \bibinfo{eid}{230802}
  (\bibinfo{year}{2012}{\natexlab{b}}).

\bibitem[{\citenamefont{{Flambaum} and {Ginges}}(2005)}]{FlaGin05}
\bibinfo{author}{\bibfnamefont{V.~V.} \bibnamefont{{Flambaum}}}
  \bibnamefont{and} \bibinfo{author}{\bibfnamefont{J.~S.}
  \bibnamefont{{Ginges}}}, \bibinfo{journal}{\pra}
  \textbf{\bibinfo{volume}{72}}, \bibinfo{eid}{052115} (\bibinfo{year}{2005}).

\bibitem[{EPA()}]{EPAPS}
\bibinfo{note}{See Supplemental Material at [URL] for illustration of the size
  of various contribution to the energies of Ag-like Nd$^{13+}$, In-like
  Pr$^{10+}$, and Sn-like Pr$^{9+}$ ions.}

\bibitem[{\citenamefont{Sugar and Kaufman}(1981)}]{ag-like-81}
\bibinfo{author}{\bibfnamefont{J.}~\bibnamefont{Sugar}} \bibnamefont{and}
  \bibinfo{author}{\bibfnamefont{V.}~\bibnamefont{Kaufman}},
  \bibinfo{journal}{Phys. Scr.} \textbf{\bibinfo{volume}{24}},
  \bibinfo{pages}{742} (\bibinfo{year}{1981}).

\bibitem[{\citenamefont{Joshi et~al.}(2001)\citenamefont{Joshi, Ryabtsev, and
  Churilov}}]{in-like-01}
\bibinfo{author}{\bibfnamefont{Y.~N.} \bibnamefont{Joshi}},
  \bibinfo{author}{\bibfnamefont{A.~N.} \bibnamefont{Ryabtsev}},
  \bibnamefont{and} \bibinfo{author}{\bibfnamefont{S.~S.}
  \bibnamefont{Churilov}}, \bibinfo{journal}{Phys. Scr.}
  \textbf{\bibinfo{volume}{64}}, \bibinfo{pages}{326} (\bibinfo{year}{2001}).

\bibitem[{nis()}]{nist-web}
\bibinfo{note}{Yu. Ralchenko, F.-C. Jou, D.E. Kelleher, A.E. Kramida, A.
  Musgrove, J. Reader, W.L. Wiese, and K. Olsen (2005). NIST Atomic Spectra
  Database (version 3.0.2), [Online]. Available: http://physics.nist.gov/asd3
  [2006, January 4]. National Institute of Standards and Technology,
  Gaithersburg, MD.}

\bibitem[{Moh()}]{MohTayNew11}
\bibinfo{note}{P.J. Mohr, B.N. Taylor, and D.B. Newell (2011), "The 2010 CODATA
  Recommended Values of the Fundamental Physical Constants" (Web Version 6.0).
  This database was developed by J. Baker, M. Douma, and S. Kotochigova.
  Available: http://physics.nist.gov/constants. National Institute of Standards
  and Technology, Gaithersburg, MD 20899.}

\bibitem[{\citenamefont{{Dzuba} and {Flambaum}}(2008)}]{DzuFla08}
\bibinfo{author}{\bibfnamefont{V.~A.} \bibnamefont{{Dzuba}}} \bibnamefont{and}
  \bibinfo{author}{\bibfnamefont{V.~V.} \bibnamefont{{Flambaum}}},
  \bibinfo{journal}{Phys. Rev. A} \textbf{\bibinfo{volume}{77}},
  \bibinfo{eid}{012515} (\bibinfo{year}{2008}).

\end{thebibliography}
\end{document}